\documentclass[twocolumn,showpacs,amsmath,amssymb,pra]{revtex4}

\usepackage{graphicx}   
\usepackage{dcolumn}    
\usepackage{bm}         
\newcommand{\rd}{{\rm d}}
\newcommand{\re}{{\rm e}}
\newcommand{\ri}{{\rm i}}
\newcommand{\kL}{k_{\rm L}}
\newcommand{\Er}{E_{\rm r}}
\newcommand{\Jz}{{\rm J}_0}

\begin{document}

\title[Dynamic localization]
      {Exploring dynamic localization with a Bose-Einstein condensate}

\author{Andr\'e Eckardt$^1$}
\author{Martin Holthaus$^2$}

\affiliation{$^1$ICFO-Institut de Ci\`{e}ncies Fot\`{o}niques,
    E-08860 Castelldefels (Barcelona), Spain}
\affiliation{$^2$Institut f\"ur Physik, Carl von Ossietzky
Universit\"at,
    D-26111 Oldenburg, Germany}

\author{Hans Lignier$^{3,4}$}
\author{Alessandro Zenesini$^{3,5}$}
\author{Donatella Ciampini$^{3,5}$}
\author{Oliver Morsch$^{3,4}$}
\author{Ennio Arimondo$^{3,4,5}$}

\affiliation{$^3$Dipartimento di Fisica ``E.~Fermi'',
    Universit\`a di Pisa, Largo Pontecorvo 3,
    56127 Pisa, Italy}
\affiliation{$^4$CNR-INFM, Largo Pontecorvo 3,
    56127 Pisa, Italy}
\affiliation{$^5$CNISM UdR Universit\`a di Pisa, Largo Pontecorvo 3,
    56127 Pisa, Italy}

\date{December 1, 2008}

\begin{abstract}
We report on the experimental observation of dynamic localization of a
Bose-Einstein condensate in a shaken optical lattice, both for sinusoidal
and square-wave forcing. The formulation of this effect in terms of a
quasienergy band collapse, backed by the excellent agreement of the
observed collapse points with the theoretical predictions, suggests
the feasibility of systematic quasienergy band engineering.
\end{abstract}


\maketitle


\section{Introduction}

The seminal work by Dunlap and Kenkre~\cite{DunlapKenkre86} has led to the
recognition that an external uniform ac force can control the spreading
of the wave packet of a particle moving in a spatially periodic potential.
In particular, an initially localized wave packet of a monochromatically
driven particle on a nearest-neighbor tight-binding lattice remains perpetually
localized for certain values of the driving amplitude~\cite{DunlapKenkre86}.
This phenomenon, termed ``dynamic localization'', is a quantum mechanical
manifestation of the fact that a time-periodic force sometimes can
stabilize a system, as shown in classical mechanics by the example of the
driven, inverted pendulum~\cite{Butikov01}. Dynamic localization is closely
related to the coherent destruction of tunneling experienced by a single
particle in a double-well potential under the influence of an ac
force~\cite{GrossmannEtAl91,LlorentePlata92,GrifoniHanggi98,KayanumaSaito08},
and to the modification of atomic $g$-factors in oscillating magnetic
fields~\cite{HarocheEtAl70,HolthausHone96}. In view of possible applications
to electrons in terahertz-driven semiconductor superlattices, it has been
shown theoretically that dynamic localization survives in the presence of
Coulomb interactions~\cite{MeierEtAl95}. However, experiments with such
devices are difficult to perform, and have to cope with a host of competing
effects~\cite{KeayEtAl95}. While there exist clean visualizations of dynamic
localization in optical analogs of driven quantum systems, realized by means
of curved waveguide arrays for light~\cite{LonghiEtAl06,DellaValleEtAl07},
the observation of dynamic localization of matter waves has long remained a
challenge.

The situation changed with the availability of ultracold atoms in optical
potentials. In 1998 Madison {\em et al.\/} obtained evidence for band
narrowing with cold sodium atoms in a phase-modulated optical lattice,
in good agreement with theoretical calculations going beyond both the
tight-binding and the single-band approximation~\cite{MadisonEtAl98}.
More recently, the basic mechanism responsible for dynamic localization,
an effective rescaling of the hopping matrix element induced by the ac force,
has been observed for single-particle tunneling in strongly driven double-well
potentials~\cite{KierigEtAl08}. It had also been pointed out that this
mechanism remains effective even for an interacting Bose-Einstein condensate,
at least for sufficiently high driving frequencies~\cite{EckardtEtAl05}. This
has led to the experimental demonstration of dynamic control of matter-wave
tunneling in an optical lattice~\cite{LignierEtAl07}, and to the observation
of photon-assisted tunneling of a condensate~\cite{SiasEtAl08}. Since the
reduction of interwell tunneling increases the relative importance of the
particles' repulsion, even the superfluid-to-Mott insulator transition of
ultracold atoms in an optical lattice can be induced by shaking the lattice
in a time-periodic manner, as has now been shown by Zenesini
{\em et al.\/}~\cite{ZenesiniEtAl08}

In the present paper we take up this line of investigation and explore dynamic
localization of a dilute Bose-Einstein condensate in a time-periodically
shifted optical lattice in more detail. We proceed as follows: In
Sec.~\ref{S_2} we cast the concept of single-particle dynamic localization
into a form that lends itself in a particularly transparent manner to further
generalizations, relying on the idea of quasienergy bands~\cite{HolthausHone93}.
We then apply these general considerations in Sec.~\ref{S_3} to optical cosine
lattices, taking into account all couplings that are omitted in the
nearest-neighbor approximation. In Sec.~\ref{S_4} we report our experimental
results, achieved with condensates of~$^{87}$Rb. Besides sinusoidal forcing,
we also consider the case of square-wave forcing, which is known to produce
exact single-particle dynamic localization for any form of the energy
dispersion~\cite{ZhuEtAl99,DignamSterke02}. The final Sec.~\ref{S_5}
contains our conclusions.

\section{The principle underlying dynamic localization}
\label{S_2}

We consider a time-dependent Hamiltonian
\begin{equation}
    H(t) = H_0 + H_1(t) \; ,
\label{eq:HAM}
\end{equation}
where
\begin{equation}
    H_0 = \frac{p^2}{2m} + V(x)
\end{equation}
describes a single particle with mass~$m$ moving in a one-dimensional periodic
potential $V(x) = V(x+d)$ with lattice constant~$d$, and
\begin{equation}
    H_1(t) = -F(t)x
\label{eq:HFO}
\end{equation}
introduces an external force $F(t)$, mediated by the position operator~$x$
in a manner analogous to electromagnetic forces on atoms within the dipole
approximation. The energy eigenfunctions $\varphi_{n,k}(x)$ of $H_0$ are
Bloch states with band index $n$ and wave number $k$,
\begin{equation}
    \varphi_{n,k}(x) = \re^{\ri k x} v_{n,k}(x) \; ,
\end{equation}
with functions $v_{n,k}(x) = v_{n,k}(x+d)$ sharing the spatial lattice
period~$d$. The solutions to the eigenvalue equation
\begin{equation}
    H_0 \varphi_{n,k}(x) = E_n(k) \varphi_{n,k}(x)
\end{equation}
provide the dispersion relations $E_n(k)$ for the various energy bands.
It is of interest to observe that the external forcing does not truly
break translational symmetry: Introducing the unitary transfomation
\begin{equation}
    U = \exp\!\left(-\frac{\ri}{\hbar}\int_{t_0}^t \! \rd \tau \,
    F(\tau) x \right) \; ,
\label{eq:UTR}
\end{equation}
the transformed wave functions $\widetilde{\psi}(x,t) = U \psi(x,t)$ are
governed by the Hamiltonian
\begin{equation}
    \widetilde{H}(t) =
    \frac{1}{2m} \left( p + \int_{t_0}^t \! \rd \tau \, F(\tau) \right)^2
    + V(x) \; ,
\label{eq:HTR}
\end{equation}
which still remains invariant under spatial translations by~$d$.

In the following, we assume band gaps so wide that, despite the external
force, the dynamics can be restricted to the lowest band. Dropping the
band index, we expand the Bloch waves of that band with respect to the
corresponding Wannier states~\cite{Wannier37,Kohn59},
\begin{equation}
    \varphi_{k}(x) = \sum_{\ell} w_\ell(x) \re^{\ri k \ell d} \; ,
\label{eq:BLW}
\end{equation}
where $w_\ell(x) = w_0(x - \ell d)$ denotes a Wannier function centered
around the $\ell$-th lattice site. We take the potential to be symmetric,
$V(x) = V(-x)$, so that $w_0(x)$ can be chosen real and symmetric~\cite{Kohn59}.
Defining the time-dependent wave number
\begin{equation}
    q_k(t) = k + \frac{1}{\hbar}\int_{t_0}^t \! \rd \tau \, F(\tau) \; ,
\label{eq:WNQ}
\end{equation}
obeying $\hbar \dot{q}_k(t) = F(t)$ and containing a still unspecified
lower integration bound~$t_0$, the wave functions
\begin{equation}
    \psi_k(x,t) = \sum_\ell w_\ell(x) \re^{\ri q_k(t) \ell d}
    \exp\!\left(-\frac{\ri}{\hbar}\int_0^t\! \rd \tau \,
    E\big(q_k(\tau)\big)\right)
\label{eq:HOU}
\end{equation}
then are ``weak'' solutions to the time-dependent Schr\"odinger equation
with the full Hamiltonian~(\ref{eq:HAM}), in the following sense: By
construction, one has
\begin{eqnarray}
    & &
    \ri\hbar\frac{\partial}{\partial t} \psi_k(x,t)
    = E\big( q_k(t)\big) \psi_k(t)
\\ & &
    - F(t) \sum_\ell \ell d \, w_\ell(x) \re^{\ri q_k(t) \ell d}
    \exp\!\left(-\frac{\ri}{\hbar}\int_0^t\! \rd \tau \,
    E\big( q_k(\tau)\big)\right) \, .
\nonumber
\end{eqnarray}
If we stipulate that the lattice site labeled $\ell = 0$ is situated
at $x = 0$, implying
\begin{equation}
    \langle 0 \, | \, x \, | \, 0 \rangle
    = \int \! \rd x \, x | w_0(x) |^2
    = 0 \; ,
\end{equation}
then one has
\begin{equation}
    \langle \ell \, | \, x \, | \, m \rangle
    = \left\{ \begin{array}{cl}
    \ell d \; & , \quad \text{if} \;\; \ell = m \\
    0      \; & , \quad \text{else}
    \end{array} \right. \; ,
\end{equation}
where $| \ell \rangle$ is the Dirac ket corresponding to $w_\ell(x)$. Note
that this is an identity, which holds exactly even for shallow lattices.
Hence, it follows that
\begin{eqnarray}
    \langle \ell \, | \,
    \ri\hbar \frac{\partial}{\partial t} \psi_k(x,t) \rangle
    & = &
    \langle \ell \, | \, E\big(q_k(t)\big) - F(t) x \, | \,
    \psi_k(x,t) \rangle
\nonumber \\ & = &
    \langle \ell \, | \, H_0 + H_1(t) \, | \, \psi_k(x,t) \rangle
\end{eqnarray}
for each~$\ell$.

The above wave functions~(\ref{eq:HOU}) can be regarded as Houston
states~\cite{Houston39}, also known as ``accelerated Bloch states'', and apply
to any type of uniform forcing~$F(t)$, provided the single-band approximation
remains viable. If we now require that the forcing be periodic with period~$T$
and zero average, so that $F(t) = F(t+T)$ and
\begin{equation}
    \frac{1}{T} \int_0^T \! \rd t \, F(t) = 0 \; ,
\end{equation}
a further step can be made: In this case the transformed
Hamiltonian~(\ref{eq:HTR}) is periodic in both space and time, thus giving
rise to spatio-temporal Bloch waves
\begin{equation}
    \widetilde{\psi}_k(x,t) = \widetilde{v}_k(x,t)
    \exp\big(\ri k x - \ri \varepsilon(k) t/\hbar\big)
\label{eq:STB}
\end{equation}
with quasimomenta $\hbar k$, quasienergies $\varepsilon(k)$,
and time-dependent Bloch functions obeying
$\widetilde{v}(x,t) = \widetilde{v}(x+d,t) = \widetilde{v}(x,t+T)$.

When deriving these states within the single-band approximation from
the Houston states~(\ref{eq:HOU}), a slight subtlety comes into play:
The lower integration bound~$t_0$ in the definition~(\ref{eq:WNQ}) of
$q_k(t)$ effectuates a shift of the wave number~$k$. In order to avoid
this shift, and thus to make sure that the wave number~$k$ labeling a
spatio-temporal Bloch wave is the same as the one which labels the Bloch
state continuously connected to it when the driving force vanishes,
this lower bound $t_0$ has to be chosen such that the integrated force
also has zero average, requiring
\begin{equation}
    \frac{1}{T} \int_0^T \! \rd t \, \int_{t_0}^t \! \rd \tau \,
    F(\tau) = 0 \; .
\label{eq:REQ}
\end{equation}
With this specification of $q_k(t)$, we set
\begin{equation}
    \varepsilon(k) = \frac{1}{T} \int_0^T \! \rd t \,
    E\big(q_k(t)\big) \; ,
\label{eq:EPS}
\end{equation}
and write the Houston states~(\ref{eq:HOU}) in the form
\begin{equation}
    \psi_k(x,t) = u_k(x,t)
    \exp\!\left(-\frac{\ri}{\hbar}\varepsilon(k) t \right) \; .
\label{eq:FLO}
\end{equation}
This formal-looking manipulation is of key significance: After the average
phase growth has been factored out, the remaining functions $u_k(x,t)$
acquire the temporal periodicity of the force,
\begin{eqnarray}
    u_k(x,t) & = & \sum_\ell w_\ell(x) \re^{\ri q_k(t) \ell d}
\nonumber \\ & & \times
    \exp\!\left(-\frac{\ri}{\hbar}\int_0^t\! \rd \tau \,
    \Big(E\big(q_k(\tau)\big) - \varepsilon(k)\Big) \right)
\nonumber \\ & = &
    u_k(x,t+T) \; .
\end{eqnarray}
Thus, these wave functions~(\ref{eq:FLO}) are Floquet
states~\cite{HolthausHone96,HolthausHone93}, from which the spatio-temporal
Bloch waves~(\ref{eq:STB}) are obtained by applying the unitary
transformation~(\ref{eq:UTR}). At times $t = t_0 + sT$ (with integer~$s$)
they coincide with the Bloch waves~(\ref{eq:BLW}), apart from a phase factor.
{\em Any\/} single-band wave packet $\psi(x,t)$ can be expanded with respect
to these basis states,
\begin{equation}
    \psi(x,t) = \left(\frac{d}{2\pi}\right)^{1/2} \!\!
    \int_{-\pi/d}^{\pi/d} \! \rd k \, a(k) u_k(x,t)
        \exp\!\left(-\frac{\ri}{\hbar}\varepsilon(k) t \right) \, ,
\label{eq:WAP}
\end{equation}
with some density $a(k)$ which is time-independent despite the presence of
the ac force. The $T$-periodic nature of the functions $u_k(x,t)$ directly
reflects the short-term response of such a packet to the force~$F(t)$, whereas
the quasienergy dispersion $\varepsilon(k)$ governs its long-time averaged
evolution in the same manner as the original energy dispersion $E(k)$
governs its evolution when $F(t) = 0$. For example, the average velocity
$\bar{v}_{k_0}$ of a wave packet initially centered around some wave
number~$k_0$ is given, in the presence of the
forcing, by
\begin{equation}
    \bar{v}_{k_0} = \frac{1}{\hbar}
    \left. \frac{\partial \varepsilon(k)}{\partial k} \right|_{k=k_0} \; .
\end{equation}
Moreover, the representation~(\ref{eq:WAP}) directly leads to a quite general
and intuitive condition for dynamic localization: Exact dynamic localization
obviously occurs when the quasienergy band collapses, so that all quasienergies
coincide, $\varepsilon(k) = \varepsilon_0$ for all $k$, implying that the
effective mass of a driven Bloch particle becomes infinite. All Floquet states
comprising the wave packet~(\ref{eq:WAP}) then acquire exactly the same phase
in the course of one driving period, since
\begin{equation}
    \psi(x,t) =
    \exp\!\left(-\frac{\ri}{\hbar}\varepsilon_0 t \right)
    \left(\frac{d}{2\pi}\right)^{1/2} \!\!
    \int_{-\pi/d}^{\pi/d} \! \rd k \, a(k) u_k(x,t) \; ,
\end{equation}
so that the packet can neither move (apart from its residual $T$-periodic
motion) nor spread, but simply reproduces itself $T$-periodically, regardless
of the form of the density $a(k)$. When this condition of zero width of the
quasienergy band cannot be met exactly, but there still is significant band
narrowing, wave packet spreading can at least be reduced substantially by
the external force~\cite{HolthausHone93}.

\section{Application to optical lattices}
\label{S_3}

For a lattice with inversion symmetry, the original energy dispersion
takes the form
\begin{equation}
    E(k) = E_0 + \sum_{\ell=1}^{\infty}
    2 \langle 0 \, | \, H_0 \, | \, \ell \rangle \cos(\ell kd) \; ,
\label{eq:DIS}
\end{equation}
with coefficients given by matrix elements of $H_0$ between Wannier states
located $\ell$~sites apart from each other. In particular, for an optical
cosine lattice of the
form~\cite{MorschOberthaler06}
\begin{equation}
    V(x) = \frac{V_0}{2}\cos(2\kL x) \; ,
\label{eq:OLA}
\end{equation}
where $\kL$ is the wave number of the lattice-generating laser radiation
and the depth $V_0$ is proportional to the intensity of that
radiation~\cite{MorschOberthaler06}, the Wannier states can easily be
computed numerically~\cite{Slater52,BoersEtAl07}. Table~\ref{T_1} lists the
resulting coupling coefficients $\langle 0 | H_0 | \ell \rangle$ for the
lowest band as functions of the scaled lattice depth $V_0/\Er$, with
\begin{equation}
    \Er = \frac{\hbar^2 \kL^2}{2m}
\end{equation}
denoting the single-photon recoil energy of an atom with mass~$m$. The
relative error committed when adopting the commonly used nearest-neighbor
approximation, {\em i.e.\/}, when neglecting all
$\langle 0 | H_0 | \ell \rangle$ with $\ell \ge 2$, is on the order of 10\%
for an optical lattice with $V_0/\Er = 3$, and reduces to 1\% only when
$V_0/\Er \approx 10$~\cite{BoersEtAl07}.

\begin{table}
\begin{tabular}{||c|c|c|c|c||} \hline
$~V_0/E_{\rm r}~$  &  $c_1$  &  $c_2$       & $c_3$                 & $c_4$           \\ \hline
$ 2.0$ &  $-0.14276$ & $2.04 \cdot 10^{-2}$ & $-4.83 \cdot 10^{-3}$ & $1.40 \cdot 10^{-3}$ \\
$ 4.0$ &  $-0.08549$ & $6.15 \cdot 10^{-3}$ & $-7.16 \cdot 10^{-4}$ & $1.01 \cdot 10^{-4}$ \\
$ 6.0$ &  $-0.05077$ & $1.91 \cdot 10^{-3}$ & $-1.15 \cdot 10^{-4}$ & $8.31 \cdot 10^{-6}$ \\
$ 8.0$ &  $-0.03080$ & $6.35 \cdot 10^{-4}$ & $-2.08 \cdot 10^{-5}$ & $8.20 \cdot 10^{-7}$ \\
$10.0$ &  $-0.01918$ & $2.27 \cdot 10^{-4}$ & $-4.25 \cdot 10^{-6}$ & $9.57 \cdot 10^{-8}$ \\
$12.0$ &  $-0.01225$ & $8.66 \cdot 10^{-5}$ & $-9.65 \cdot 10^{-7}$ & $1.29 \cdot 10^{-8}$ \\
$14.0$ &  $-0.00800$ & $3.49 \cdot 10^{-5}$ & $-2.39 \cdot 10^{-7}$ & $1.96 \cdot 10^{-9}$ \\
$16.0$ &  $-0.00533$ & $1.47 \cdot 10^{-5}$ & $-6.37 \cdot 10^{-8}$ & $3.31 \cdot 10^{-10}$ \\
$18.0$ &  $-0.00362$ & $6.48 \cdot 10^{-6}$ & $-1.81 \cdot 10^{-8}$ & $6.09 \cdot 10^{-11}$ \\
$20.0$ &  $-0.00249$ & $2.95 \cdot 10^{-6}$ & $-5.45 \cdot 10^{-9}$ & $1.22 \cdot 10^{-11}$ \\ \hline
\end{tabular}
\caption{Hopping matrix elements
    $c_\ell = \langle 0 | H_0 | \ell \rangle/\Er$ for the lowest band of
    an optical cosine lattice~(\ref{eq:OLA}), as functions of the scaled
    lattice depth $V_0/E_{\rm r}$. Observe that the sign of $c_\ell$
    alternates with $\ell$.}
\label{T_1}
\end{table}

\subsection{Sinusoidal forcing}

\begin{figure}[t]
\includegraphics[angle=-90., scale=0.5, width = 7cm]{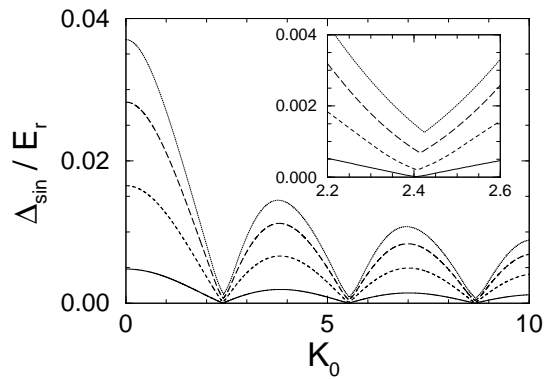}
\caption{Width $\Delta_{\rm sin}$ of the lowest quasienergy band of an
    optical cosine lattice under sinusoidal driving as function of the
    dimensionless amplitude~(\ref{eq:SDA}), for $V_0/\Er = 2$ (dots),
    $3$ (long dashes), $5$ (short dashes), and $10$ (full line).
    The inset quantifies the extent of band narrowing for values of
    $K_0$ close to the first zero $j_{0,1} = 2.405$ of the Bessel
    function ${\rm J}_0$.}
\label{F_1}
\end{figure}

We now consider purely sinusoidal forcing, as specified by
$F(t) = F_0 \cos(\omega t + \varphi)$ with some arbitrary phase~$\varphi$.
In this case, Eq.~(\ref{eq:WNQ}) together with the requirement~(\ref{eq:REQ})
yields
\begin{equation}
    q_k(t) = k + \frac{F_0}{\hbar\omega}\sin(\omega t + \varphi) \; ,
\end{equation}
implying
\begin{equation}
    \frac{1}{T} \int_0^T \! \rd t \, \cos \big( q_k(t) \ell d \big)
    = \Jz(\ell K_0) \, \cos(\ell k d) \; ,
\end{equation}
where $\Jz(z)$ is a Bessel function of order zero, and we have introduced
the scaled driving amplitude
\begin{equation}
    K_0 = \frac{F_0 d}{\hbar\omega} \; .
\label{eq:SDA}
\end{equation}
Hence, the quasienergy dispersion resulting from the Bloch band~(\ref{eq:DIS})
under sinusoidal forcing reads
\begin{equation}
    \varepsilon(k) = E_0 + \sum_{\ell=1}^{\infty}
    2 \langle 0 \, | \, H_0 \, | \, \ell \rangle \,
    \Jz(\ell K_0) \, \cos(\ell kd) \; .
\label{eq:ESN}
\end{equation}
For the particular example of quasienergy bands originating from the lowest
Bloch band of an optical lattice~(\ref{eq:OLA}), characterized by the matrix
elements collected in Table~\ref{T_1}, the quasienergy band widths are
depicted in Fig.~\ref{F_1} as functions of $K_0$ for some typical depths~$V_0$.
Due to the dominance of the nearest-neighbor hopping matrix element
$\langle 0 | H_0 | 1 \rangle$ there is strong band narrowing for values
of $K_0$ close to the zeros of the Bessel function ${\rm J}_0$, but the
nonvanishing longer-range hopping elements prevent the band from collapsing
completely, as emphasized by the inset. Thus, one expects appreciable, but
incomplete dynamic localization in sinusoidally driven shallow optical
lattices.

\subsection{Square-wave forcing}

\begin{figure}[t]
\includegraphics[angle=-90., scale=0.5, width = 7cm]{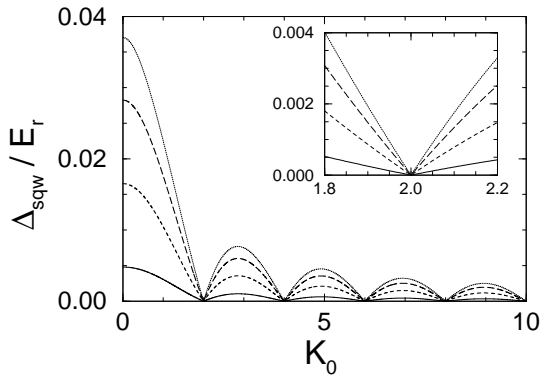}
\caption{Width $\Delta_{\rm sqw}$ of the lowest quasienergy band of an
    optical cosine lattice under square-wave driving as function of the
    dimensionless amplitude~(\ref{eq:SDA}), for $V_0/\Er = 2$ (dots),
    $3$ (long dashes), $5$ (short dashes), and $10$ (full line).
    The inset illustrates that an exact band collapse occurs when
    $K_0$ is a nonzero integer multiple of~$2$.}
\label{F_2}
\end{figure}

For square-wave forcing of the form
\begin{equation}
    F(t) = \left\{ \begin{array}{rl}
        F_0 \; & , \quad 0 \leq t < T/2 \\
           -F_0 \; & , \quad T/2 \leq t < T \; ,
           \end{array} \right.
\end{equation}
Eqs.~(\ref{eq:WNQ}) and~(\ref{eq:REQ}) yield
\begin{equation}
    q_k(t) = \left\{ \begin{array}{ll}
        k + F_0 (t-T/4)/\hbar   \; & , \quad 0 \leq t < T/2 \\
            k + F_0(3T/4-t)/\hbar   \; & , \quad T/2 \leq t < T
           \end{array} \right. \; ,
\end{equation}
to be continued $T$-periodically to all~$t$. This gives
\begin{equation}
    \frac{1}{T}\int_0^T \! \rd t \, \cos \big(q_k(t) \ell d \big)
    = {\rm sinc}\!\left(\frac{\ell\pi K_0}{2}\right)
    \cos(\ell kd) \; ,
\end{equation}
with $K_0$ being defined according to Eq.~(\ref{eq:SDA}), having set
$\omega = 2\pi/T$. Moreover, we write ${\rm sinc}(z) = \sin(z)/z$.
Correspondingly, the quasienergy dispersion becomes
\begin{equation}
    \varepsilon(k) = E_0 + \sum_{\ell=1}^{\infty}
    2 \langle 0 \, | \, H_0 \, | \, \ell \rangle \,
    {\rm sinc}\!\left(\frac{\ell\pi K_0}{2}\right)
    \cos(\ell kd) \; .
\label{eq:ESW}
\end{equation}
Since all ${\rm sinc}$-functions adopt their zeros simultaneously, there
is a total collapse of this quasienergy band when $K_0 = 2\nu$ with integer
$\nu = 1,2,3,\ldots \; $. Thus, we recover the known fact that there is exact
dynamic localization, regardless of both the values of the hopping matrix
elements and the form of the wave packet, for these particular driving
amplitudes~\cite{ZhuEtAl99,DignamSterke02}. Figure~\ref{F_2} shows
the widths of quasienergy bands originating from the lowest Bloch band
of an optical lattice under square-wave forcing; the total band collapse
at $K_0 = 2$ is clearly visible in the inset.

\section{Experimental results}
\label{S_4}

Our experimental setup is described in detail in Ref.~\cite{LignierEtAl07}. 
Briefly, we adiabatically load Bose-Einstein condensates consisting of about 
$6 \times 10^{4}$ atoms of $^{87}$Rb into the lowest band of a one-dimensional
optical lattice. The lattice is generated by focusing two counter-propagating 
linearly polarized laser beams of wavelength $\lambda = 2\pi/\kL = 842$~nm 
onto the condensate, resulting in the periodic potential~(\ref{eq:OLA}) 
along the beam direction. Each beam passes through an acousto-optic modulator, 
allowing us to introduce a frequency difference $\Delta \nu(t)$ between the
beams that can be used to accelerate or shake the lattice. In the laboratory 
frame of reference, the condensate then experiences a time-dependent potential
\begin{equation}
    V_{\rm lab}(x,t) = \frac{V_0}{2}\cos\!\left(2\kL \left[ x -
    \frac{\lambda}{2} \int_{t_1}^t \! \rd \tau \, \Delta\nu(\tau)
    \right]\right) \; .
\end{equation}
By means of a unitary transformation to the comoving frame of reference, the
single-particle Hamiltonian is brought into the form
\begin{equation}
    H = \frac{p^2}{2m} + \frac{V_0}{2}\cos(2\kL x)
    + m \frac{\lambda}{2} \frac{\rd \, \Delta \nu(t)}{\rd t} \, x \; .
\end{equation}
Hence, prescribing an oscillating frequency difference
\begin{equation}
    \Delta \nu(t) = \Delta \nu_{\rm max} \sin(\omega t)
\end{equation}
with amplitude $\Delta \nu_{\rm max}$, one obtains sinusoidal forcing with
strength $F_0 = m\omega \lambda \Delta \nu_{\rm max}/2$, amounting to the
scaled driving amplitude
\begin{equation}
    K_0 = \frac{\pi^2}{2} \, \frac{\hbar\Delta\nu_{\rm max}}{\Er} \; .
\end{equation}
On the other hand, the triangular protocol
\begin{equation}
    \Delta\nu(t) = \left\{ \begin{array}{ll}
    2 \Delta\nu_{\rm max} t/T   \; & , \quad 0 \leq t < T/2 \\
    2 \Delta\nu_{\rm max} (1 - t/T) \; & , \quad T/2 \leq t < T
    \end{array} \right.
\end{equation}
results in square-wave forcing with strength
$F_0 = m\lambda \Delta\nu_{\rm max}/T$, giving
\begin{equation}
    K_0 = \frac{\pi}{2} \, \frac{\hbar\Delta\nu_{\rm max}}{\Er} \; .
\end{equation}

We first study the {\em in situ\/} expansion of a condensate under both
kinds of forcing~\cite{LignierEtAl07}. To this end, initially an anisotropic,
elongated harmonic trap with trapping frequencies of 20~Hz (longitudinally)
and 80~Hz (radially) was superimposed on the lattice. After switching off
the laser beam effectuating the longitudinal confinement, the condensate was
free to expand in the direction of the shaken lattice. After some time, it
was illuminated by a resonant flash, the shadow cast by which was imaged
onto a CCD chip.

We work with lattices possessing a fairly large depth between $V_0 = 6 \, \Er$
and $V_0 = 9 \, \Er$, so that the effect of longer-range hopping is not
discernible at the scale of our experimental accuracy, and we may employ
the nearest-neighbor approximation, keeping only the hopping matrix element
$J = -\langle 0 |H_0 | 1 \rangle$ connecting adjacent lattice sites. Thus,
the unperturbed energy band is well described by the cosine dispersion
relation
\begin{equation}
    E(k) = -2J \cos(kd) \; ,
\end{equation}
assuming that interparticle interactions remain negligible. Accordingly,
the quasienergy band of the driven system is approximated by
\begin{equation}
    \varepsilon(k) = -2J_{\rm eff} \cos(kd) \; ,
\end{equation}
where the effective hopping matrix element $J_{\rm eff}$ is gven by
\begin{equation}
    J_{\rm eff} = J \, {\rm J_0}(K_0)
\label{eq:JSN}
\end{equation}
for sinusoidal forcing, as follows from Eq.~(\ref{eq:ESN}), whereas
Eq.~(\ref{eq:ESW}) gives
\begin{equation}
    J_{\rm eff} = J \, {\rm sinc}(\pi K_0/2)
\label{eq:JSW}
\end{equation}
for a square-wave drive. The measured expansion rate $\rd \sigma_x/ \rd t$ 
of the condensate width $\sigma_x$ along the lattice direction is then to 
a good approximation proportional to the hopping element $J_{\rm eff}$ 
which effectively describes nearest-neighbor tunneling when the forcing is
present~\cite{LignierEtAl07}. This is a working example of our general 
philosophy: The time-averaged evolution of the driven system proceeds in 
close analogy to that of an undriven one, with $\varepsilon(k)$ replacing 
$E(k)$. In practice, rather than extracting $J_{\rm eff}$ from the expansion 
data, for each experiment we also measure the bare expansion rate (which is
proportional to $J$) in the undriven system, from which the normalized 
effective hopping element $J_{\rm eff}/J$ is then calculated.

\begin{figure}[t]
\includegraphics[angle=0., scale=0.5, width = 7cm]{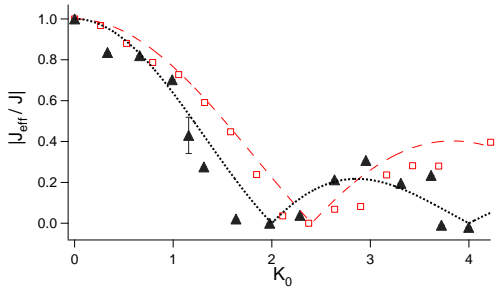}
\caption{(Color online) Absolute value of the effective hopping matrix
    element $J_\mathrm{eff}$ for a condensate in an optical lattice with
    depth $V_0/\Er = 6$ under the influence of sinusoidal (open boxes) and
    square-wave (triangles) driving, normalized to the ``bare'' hopping
    element~$J$ which governs the condensate spreading in the undriven
    lattice. The lines correspond to the approximations~(\ref{eq:JSN})
    and~(\ref{eq:JSW}). The driving frequencies $\omega/2\pi$ were 
    $1.0\,\mathrm{kHz}$ for sinusoidal driving, and $1.5\,\mathrm{kHz}$ 
    for the square-wave case.}
\label{F_3}
\end{figure}

In Fig.~\ref{F_3} we plot our results for both types of forcing versus 
the scaled driving amplitude~$K_0$, obtained with a sinusoidal drive of
frequency $\omega/2\pi = 1.0 \, \mathrm{kHz}$, and a square-wave one 
with $\omega/2\pi = 1.5 \, \mathrm{kHz}$. Evidently, the single-particle 
expectations are matched quite well, with the data obtained for square-wave 
driving scattering a bit more strongly around the prediction~(\ref{eq:JSW}) 
than those for sinusoidal driving around the graph of Eq.~(\ref{eq:JSN}).

\begin{figure}[t]
\includegraphics[angle=0., scale=0.5, width = 7cm]{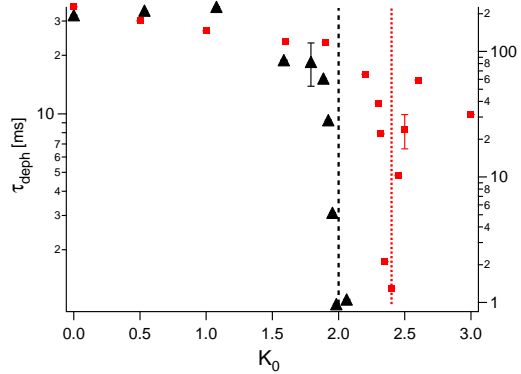}
\caption{(Color online) Dephasing times for a condensate in an optical lattice
    with depth $V_0/\Er = 9$ under the influence of square-wave (triangles;
    left scale) and sinusoidal (boxes; right scale) driving.}
\label{F_4}
\end{figure}

An alternative method for determining the points of maximum band collapse
is to measure the phase coherence of the condensate in the shaken
lattice~\cite{LignierEtAl07}. Switching off the confining potential and
letting the condensate fall under gravity for 20~ms, we obtain an interference
pattern whose visibility reflects the condensate coherence. Recording this
visibility as a function of time, we extract the decay time constant
$\tau_{\rm deph}$ of the resulting near-exponential function. In general,
we find dephasing times on the order of 100~ms in the presence of even strong
sinusoidal driving, while they are somewhat shorter for the square-wave drive,
as shown in Fig.~\ref{F_4}. In the immediate vicinity of the band collapse
points, however, the dephasing times are strongly reduced: When the quasienergy
band width approaches zero, the individual lattice sites are effectively
decoupled, so that the local phases evolve independently due to interatomic
collisions, resulting in a rapid dephasing of the array. This effect leads to
sharply pronounced dips in plots of $\tau_{\rm deph}$ vs.\ $K_0$, allowing us
to confirm the theoretically predicted first collapse points to fairly good
accuracy: $K_{0} = 2.4$ for sinusoidal forcing, whereas $K_{0} = 2.0$ for the
square-wave drive.

While in the present work we are mainly concerned with identifying the
conditions for partial or complete dynamic localization, useful insight
can also be obtained by studying the time evolution of the width of the
driven matter-wave packet away from the collapse points. In general, if
the extension $\sigma_x$ of a Floquet wave packet~(\ref{eq:WAP}) narrowly
centered in $k$-space around some $k_0$ is given by $\sigma_x = a$ at
time $t = 0$, that wave packet spreads in time according to
\begin{eqnarray}
    \sigma_x(t) & = & a \,
    \sqrt{1 + \left( \frac{\varepsilon''(k_0) t}{\hbar a^2}\right)^2}
\nonumber \\ & \sim &
    \gamma_a(k_0) t \; ,
\end{eqnarray}
with asymptotic, large-$t$ expansion rate
\begin{equation}
    \gamma_a(k_0) = \frac{| \varepsilon''(k_0) |}{\hbar a}
\end{equation}
determined by the second derivative of the quasienergy dispersion at $k_0$.
According to Eqs.~(\ref{eq:ESN}) and (\ref{eq:ESW}), here the hopping elements
$\langle 0 | H_0 | \ell \rangle$ enter with weights proportional to~$\ell^2$,
each depending in a different manner on the driving amplitude. Hence, precise
measurements of the expansion rate could yield information both on the state
of the driven packet, {\em i.e.}, on the wave number $k_0$, and on the
next-to-nearest neighbor couplings. Furthermore, while we have worked with
condensate densities so low that interparticle interactions can be ignored
and the single-particle picture remains applicable, it would be interesting
to identify signatures of interparticle interactions in dynamic localization.

\section{Conclusions and outlook}
\label{S_5}

In this work we have pointed out that wave packet motion and spreading in
time-periodically forced spatially periodic potentials are governed by
quasienergy bands $\varepsilon(k)$ in a manner which exactly parallels the
description of quantum dynamics in unforced lattices in terms of their
energy bands $E(k)$, and have experimentally demonstrated dynamic
localization, {\em i.e.\/}, ``freezing'' of the dynamics due to a
quasienergy band collapse, with dilute Bose-Einstein condensates in optical
lattices subjected to either sinusoidal or square-wave driving.
Seen from a conceptual viewpoint, this experimental confirmation of the
ideas developed in the context of dynamic localization can be regarded as a
proof of principle for the notion of {\em quasienergy band engineering\/}:
Different types of forcing, applied to the same energy band, can lead to
substantially different quasienergy dispersion relations, and hence can be
employed for realizing band structures which even might not have an analog
in traditional solid-state systems.

In order to make contact with previous theoretical works~\cite{DunlapKenkre86,
GrifoniHanggi98,HolthausHone93,ZhuEtAl99,DignamSterke02},
we have restricted our investigation here to forces with parameters which
do not lead to significant coupling of several energy bands. In this case
the construction of the quasienergy dispersion merely involves taking the
average~(\ref{eq:EPS}) and thus is an easy exercise, since one is essentially
dealing with just a free Bloch particle. However, such quasienergy bands,
reflecting the existence of spatio-temporal Bloch waves~(\ref{eq:STB}),
originate from nothing more than the simultaneous presence of a spatially
periodic potential and a temporally periodic force. Accordingly, they also
exist when several unperturbed energy bands are coupled by the force; the
quasienergy bands then exhibit fairly nontrivial features~\cite{HolthausHone96}.
The exploration of these further options for band engineering which result from
deliberate interband coupling, together with the inclusion of interparticle
interaction, should open up further interesting avenues.

\begin{acknowledgments}
A.E.\ is grateful to M.\ Lewenstein for kind hospitality at
ICFO-Institut de Ci\`{e}ncies Fot\`{o}niques, and acknowledges a
Feodor Lynen research grant from the Alexander von Humboldt foundation.
Financial support by the E.U.-STREP ``OLAQUI'' and by a CNISM ``Progetto
Innesco 2007'' for the experimental part of this work is also gratefully
acknowledged. We thank J.\ Radogostowicz , C.\ Sias and Y.\ Singh for
assistance.
\end{acknowledgments}

\end{document}